 \documentclass[preprint,aps,showpacs,nofootinbib]{revtex4}
\usepackage{graphicx,amsmath,amssymb}
\usepackage{psfrag}
\usepackage{color}
\usepackage{hyperref}
\newcommand{\be}{\begin {eqnarray}}
\newcommand{\ee}{\end{eqnarray}}

\newcommand{\la}{\langle}
\newcommand{\ra}{\rangle}

\usepackage[T1]{fontenc}
\usepackage[utf8]{inputenc}
\usepackage{lmodern}
\usepackage{amsfonts}
\usepackage{color}
\newcommand{\CB}{Borici-Creutz}
\newcommand{\p}{\pi_{\Gamma}}
\newcommand{\s}{\sigma}
\newcommand{\G}{\Gamma}

\newcommand{\den}{(\s^2+\frac{\p^2}{2}+\p(C+D)+C^2+D^2)}
\newcommand{\deno}{(\frac{\p^2}{2}+\p(C+D)+C^2+D^2)}
\newcommand{\logn}{[\s^2+\frac{\p^2}{2}+\p(C+D)+C^2+D^2]}

\begin{document}
%%%%%%%%%%%%%%%%%%%
\title{  Gross-Neveu model with Borici-Creutz fermion}
\author{ J. Goswami$^1$,  D. Chakrabarti$^1$,  S. Basak$^2$ }
\affiliation{  $^1$Department of Physics,  Indian Institute of Technology Kanpur, Kanpur-208016, India\\
$^2$School of Physical Sciences, NISER, Bhubaneswar, India.}
\date{\today}
\begin{abstract}
We investigate the chiral phase structure of the Gross-Neveu model on a 2-D lattice using the Borici-Creutz  fermion action. We present a strong coupling analysis of the Gross-Neveu model and  perform a hybrid Monte Carlo simulation of the  model  with { \CB} fermions. Both analytic and lattice results show a second-order chiral phase transition.

 \end{abstract}


\maketitle

%%%%%%%%%%%%%%%%%%%%%%%%%%%%%%%%%%
\section{Introduction}
%%%%%%%%%%%%%%%%%%%%%%%%%%%%%
%%%%%%%%%%%%%%%%%%
The simulation of light fermions on a lattice is always a challenging task. There are several prescriptions to circumvent or minimize the doubling problem without spoiling the chiral symmetry on the lattice. By the no-go theorem, the minimum number of species one can have on a lattice with chiral symmetry is two, that is what is known as  a "minimally doubled" fermion. One such formulation is given by Karsten\cite{K} and Wilczek\cite{W}. Motivated by the fact that  electrons on a graphene lattice are described by a massless Dirac-like equation (quasi-relativistic Dirac equation), Creutz\cite{creutz}  proposed a four dimensional Euclidean lattice action describing  two flavors of fermion, each centered at $\pm p_\mu$ in the momentum space. The action was defined on a honeycomb or graphene lattice with  tunable parameters to control  the magnitude of $p_\mu$. Borici\cite{borici} immediately found a solution for the  parameters such that  the two flavors are located at the origin and at $(\frac{\pi}{2},\frac{\pi}{2},\frac{\pi}{2},\frac{\pi}{2})$. The chirally- invariant  Borici-Creutz (BC) action breaks hypercubic and discrete symmetries such as parity and time-reversal and thus allows non-covriant counter terms through quantum corrections \cite{bedaque}. Later on, Creutz proposed a refinement of the BC action so that  such effects can be mitigated to the point where they are manageable\cite{creutz2}. 
Both the Karsten-Wilczek and { \CB} actions break hypercubic and discrete symmetries, but to decide  which one is better than the other requires detailed studies of the two fermion formulations.
The renormalization properties of the BC fermion at  one loop in the perturbation theory have been investigated in Ref.\cite{capitani}.
 It was shown that, in the presence of  a gauge background with integer-valued topological charge, BC action satisfies the Atiya-Singer index theorem\cite{dc}. But there is not enough numerical study in the literature to suggest that the BC action is really better than the other lattice actions or  useful for QCD simulation.

In this work, we consider the Gross-Neveu model on a space-time square  lattice with BC fermion.  
Chiral and parity-broken(Aoki) phase structures of the Gross-Neveu model have been studied for Wilson  and Karsten-Wilczek fermions \cite{CKM,misumi}. A lattice simulation of the Gross-Neveu model   using  the Wilson fermion was done by Korzec et al\cite{korzec}, where the recovery of chirally invariant Gross-Neveu model from a lattice model was studied.
The semimetal-insulator phase transition on a graphene lattice with Thirring type four fermion interactions has been studied by Hands and collaborators\cite{hands} and the strong coupling analysis of  the tight-binding graphene model with Kekule distortion term has been done by Araki\cite{araki}.
 No numerical study of the phase structure with the {\CB} fermion has  been done yet.
 In this paper, we investigate the chiral phase structure of the Gross-Neveu model on the orthogonal lattice in the strong coupling analysis and show that  numerical results from the Gross-Neveu model with  BC fermion  are  in agreement  with  the analytic predictions.  For the lattice simulation, we use the hybrid Monte Carlo(HMC) method which was shown to be a better choice over Kramers Equation Monte Carlo(KMC) for Gross-Neveu model\cite{basak}.

%%%%%%%%%%%%%%%%%%%%%%%%%%%%%%%%

%\tableofcontents

%%%%%%%%%%%%%%%%%%%%%%%%%%%%%%%%%%%%%%%%%%%%%%%%%%%%% ANALYTIC PART IN 4 D %%%%%%%%%%%%%%%%%%%%%%%%%%%%%%%%%%%%%

\section{Strong coupling analysis of the {\CB} fermions}
A strong coupling analysis of the Gross-Neveu model has been done by Misumi and collaborators\cite{misumi, CKM} for Karsten-Wilczek  minimally doubled fermions. Here we follow a similar procedure to find the chiral phase structure for the  BC action. 
The free {\CB} action in 4D is written as,
  \be
  S_{BC}&=&\sum_{n}[\frac{1}{2}\sum_{\mu}\bar{\psi}_n\gamma_{\mu}(\psi_{n+\mu}-\psi_{n-\mu})-\frac{i r}{2}\sum_{\mu}\bar{\psi}_n(\Gamma-\gamma_{\mu})(2\psi_{n}-\psi_{n+\mu}-\psi_{n-\mu}) \nonumber \\
 && +ic_{3}\bar{\psi}_{n}\Gamma\psi_{n}+m\bar{\psi}_{n}\psi_{n}]\label{CB}
  \ee
where, $\Gamma=\frac{1}{2}(\gamma_{1}+\gamma_{2}+\gamma_{3}+\gamma_{4})$ and $\{\Gamma,\gamma_{\mu}\} =1$.
 We write this action by using the hopping and on-site operators as,
   \begin{equation}
  S_{BC}=\sum_{n} [ \sum_{\mu}(\bar{\psi_{n}}P_{\mu}^{+}\psi_{n+\mu}-\bar{\psi_{n}}P_{\mu}^{-}\psi_{n-\mu})+\bar{\psi_{n}}\hat{M}\psi_{n}] ,
\end{equation}
 where,  the hopping operators are defined as $P_{\mu}^{+}=\frac{\gamma_{\mu}}{2}(1-ir)+\frac{ir\Gamma}{2}$ and $ P_{\mu}^{-}=\frac{\gamma_{\mu}}{2}(1+ir)-\frac{ir\Gamma}{2}$ 
   and the onsite operator $\hat{M}=m+i(c_{3}-2r)\Gamma$.
     In the strong coupling limit the effective action %near the phase boundary  
     can  be written as \cite{misumi}
%Now the effective actions we can write near the phase boundary as,
 \begin{equation}
   S_{eff}=N_{c}\sum_{n}[\sum_{\mu}{\text Tr}{(M(n)(P_{\mu}^{+}})^{T}M(n+\hat{\mu})({P_{\mu}^{-}})^{T})+{\text Tr}( \hat{M}M(n)) -{\text Tr} (\log M(n))]
 \end{equation}  
 where $M(n)=\bar{\psi}(n)\psi(n)/{N_c}$ and the trace is over spinor indices. The above  effective action is obtained by performing the one-link  integration  of  the gauge fields and keeping only the leading term.
 %[misumi et. all]
 As the {\CB} action specifies  the $\Gamma$ direction as a special(diagonal) direction, we expect a vector condensate  ${\p}$ along with a chiral condensate  ${\s}$. So, %we here consider a condensate which has both chiral and ${\p}$ vector condensate,
 the condensate which is the vacuum expectation value of $M(n)$,  has both chiral and  vector condensates as components,
 \begin{equation}
   \la{M(n)}\ra=M_{0}={\s}I_{4}+i{\G}{\p}.
   \end{equation}
After putting this into Eq. (3) we get the effective action as,
\begin{equation}
  S_{eff}=N_{c}[4{\s}^{2}(1+r^{2})+2{\p}^{2}(1+r^{2})+4m{\s}-4(c_{3}-2r){\p}-2\log(\s^{2}+{\p}^{2})]
\end{equation} 
The saddle point solutions are obtained from the equations
\begin{equation}
 \frac{\delta S_{eff}}{\delta \sigma}=0; ~\frac{\delta S_{eff}}{\delta \pi_{\Gamma}}=0. 
\end{equation}
From the above two equations,  we get the gap equations as
\begin{eqnarray}
  &&2{\s}(1+r^{2})+m-\frac{\s}{{\s}^{2}+{\p}^{2}}=0,\label{saddle1} \\
 &&{\p}(1+r^{2})-(c_{3}-2r)-\frac{\p}{{\s}^{2}+{\p}^{2}}=0. 
\end{eqnarray}
These equations can be solved analytically only for $m=0$.
 In the limit ${\s}\to 0$, we get the chiral boundaries for massless {\CB} fermions  at,
  \begin{eqnarray}
    c_{3}-2r=\pm\sqrt{\frac{1+r^{2}}{2}} .
   \ee
 % \mbox{after putting r=1 we get the chiral boundaries at}\nonumber  \\
 For $ r=1$  the chiral boundaries are  at
  $
    \bar{c_{3}}=c_3-2=\pm{1}. $
%   after considering    $\bar{c_{3}}=c_{3}-2$.
% Again for m=0 and r=1 if we solve the gap equations we get following solutions,
We get two solutions for the condensates for $m=0$ and $r=1$ as
 \begin{eqnarray}
  \s=0 , \p=\frac{1}{4}\big(\bar{c_{3}}\pm\sqrt{8+\bar{c_3}^{2}}~ \big) ;\label{sol}  \\
   \nonumber \\
   \s=\frac{\sqrt{1-\bar{c_{3}}^{2}}}{2}, \p=-\frac{\bar{c_{3}}}{2} .
  \end{eqnarray}
  \begin{figure*}
\includegraphics[width=10 cm]{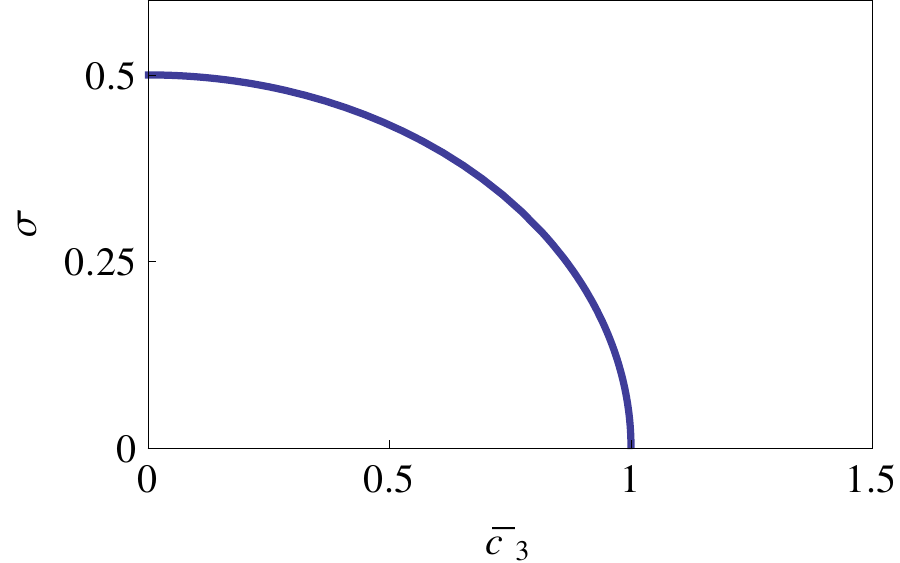}
\centering
\caption{$\bar{c_{3}}$ vs $\s$ for {\CB} fermions when $m=0$ and $r=1$.\label{ch} }
\centering
\end{figure*} 
  We see that $\s$ is only nonzero between $\bar{c_{3}}=\pm{1}$ in the massless limit. 
  Since for $\s=0$, we know from Eq.(\ref{saddle1}) that $\p=\pm 1/2$,  for $\bar{c_3}=-1$ we need to take the positive sign before the square root in the solution of $\p$ in Eq. (\ref{sol}), while for $\bar{c_3}=+1$, the negative sign should be taken.
%  As from our above solution we can see that $\bar{c_{3}}=\pm{1}$ the value of $\p=\pm{0.5}$ which is consistent with the Eq.(\ref{sol}) if we take (-) value of the $\pm\sqrt{8+\bar{c_{3}^{2}}}$.
%  Anyway, so if we plot the $\s$ for the non zero values we get the Fig.{\ref{ch}}. Which is consistent with what misumi says in his paper.
In Fig.(\ref{ch}), we have plotted the chiral condensate $\s$ as a function of the parameter $\bar{c_3}$ up to zero as  there is a discontinuity of the Dirac operator at $\bar{c_3}=0$ (there is no zero of the Dirac operator at the exact value of $\bar{c}_3=0$).
  We can clearly see the second order phase transition in agreement with Ref. \cite{misumi}.

%%%%%%%%%%%%%%%%%%%%%%%%%%%%%%%%%%%%%%% ANALYTIC PART FOR THE PHASE TRANSITION IN GROSS NEVEU MODEL %%%%%%%%%%%%%%%%%%%%%%%%%%%%%%%%%%%%%% 

\section{Phase transition for the Gross-Neveu model with {\CB} fermions}
%%%%%%%%%%%%%%%%%%%%%%%%%%%%%%%%
The {\CB}  action has already been defined in Eq.(\ref{CB}).  In 2D, $\Gamma=\frac{1}{2}(\gamma_{1}+\gamma_{2})$ and $\{\Gamma,\gamma_{\mu}\} =1$,
 and  $\Gamma^{2}=\frac{1}{2}$    and we use 2-dimensional representation for the gamma matrices. With $r=1$, 
 %%%%%%%%%%%%%%%%%%%
%  \subsection{Minimal doubling analysis}
  %%%%%%%%%%%%%%%%%
   the action describes minimally doubled fermions for $-2<c_{3}< 2$. In 2D, $$\sum_{\mu}\bar{\psi}_n(\Gamma-\gamma_\mu)\psi_n=0$$ whereas in 4D,  it produces $ 2\bar{\psi}_n \Gamma\psi_n$. We keep a similar term in 2D  by replacing  
    $c_3$ by  $(c_3-2)$  in the action.  \\
    Before proceeding further with the GN model let us first  discuss about the number of flavors of free theory with respect to the parameter $c_3$ in 2D,  flavor structure of the BC action in 4D has been studied in detail by Kimura {\it et.al.}\cite{kimura}.  
    The free Dirac operator in momentum space is  written as, 
    \be D_{BC}(p)=\sum_{\mu}[i\gamma_{\mu}\sin p_{\mu}+i(\Gamma-\gamma_{\mu})cos(p_{\mu})]+i(c_3-2)\Gamma.\ee
    Now at $c_3=0$ and $c_3=4$  the Dirac operator has only one zero with an unphysical dispersion relation,   $D_{BC}(p) \sim i\gamma_{\mu}{p_{\mu}}^2$;  hence the exact values of $c_3=0,4$ are not allowed. For ~ $0<c_3<0.59$ ~ and ~ $3.41<c_3<4$,   the Dirac operator has   two zeros with  a physical dispersion relation.  These are the two regions with minimally doubled fermion. In the rest of the region i.e  for $0.59<c_3<3.41$, the Dirac operator has four zeros. Out of those  four zeros,  the correct continuum limit of the Dirac operator is obtained  only when $p_1={p_2}$ which is satisfied by two zeros while the other two do not give the correct continuum limit i.e the dispersion relations become unphysical. For example,   for $c_3=2$, we get four zeros of the Dirac operator at  $(0,0),(\pi,\pi),(-\frac{\pi}{4},\frac{3\pi}{4}),(\frac{\pi}{4},-\frac{3\pi}{4})$, out of this only the first two zeros give physical flavors and the last two zeros give  unphysical dispersion relations. The continuum limit of the Dirac operator at $(\frac{\pi}{4},-\frac{3\pi}{4})$, looks like 
%     $$i\gamma_1[\frac{1}{\sqrt{2}}p_1+\frac{1}{2\sqrt{2}}(p_1+p_2)]+i\gamma_2[\frac{1}{\sqrt{2}}p_2-\frac{1}{2\sqrt{2}}(p_2+p_1)]$$ \\
     $\frac{1}{2\sqrt{2}}[i\gamma_1(3p_1+p_2)+i\gamma_2(p_2-p_1)],$
      so the dispersion relation  becomes unphysical for this case.  So, to study the Gross-Neveu model with a minimally doubled fermion, we need to choose a value of $c_3$ within the allowed regions i.e., $0<c_3<0.59$ or $3.41<c_3<4$. 
%%%%%%%%%%%%%%%%%%%%%%%%%%
\subsection{Gross Neveu model in 2 dimensions}
%%%%%%%%%%%%%%%%%%%%%
  The action after including the four fermion interaction (with $r=1$) is
     \begin{eqnarray}
       S_{BC}&=&\sum_{n}[\frac{1}{2}\sum_{\mu}\bar{\psi}_n\gamma_{\mu}(\psi_{n+\mu}-\psi_{n-\mu})-\frac{i}{2}\sum_{\mu}\bar{\psi}_n(\Gamma-\gamma_{\mu})(2\psi_{n}-\psi_{n+\mu}-\psi_{n-\mu})\nonumber \\
        && +i(c_{3}-2)\bar{\psi}_{n}\Gamma\psi_{n}+m\bar{\psi}_{n}\psi_{n}-\frac{g^{2}}{2N}[(\bar{\psi_{n}}\psi_{n})^{2}+(\bar\psi_{n}i\Gamma\psi_{n})^{2}],\label{gnmodel}
        \ee
where,  $g$ is the coupling constant which we consider  the same for both  four point (scalar and vector) interactions. To linearize the four-fermion interactions,  we introduce two  real auxiliary fields  $\sigma$ and $\pi_{\Gamma}$:
\begin{eqnarray*}
  \sigma(n)&=&m-\frac{g^{2}}{N}(\bar{\psi_{n}}\psi_{n}) \\
  \pi_{\Gamma}(n)&=&c_3-2-\frac{g^{2}}{N}(\bar{\psi_{n}}i\Gamma\psi_{n}).
  \end{eqnarray*}

     \begin{figure*}
\includegraphics[width =10cm]{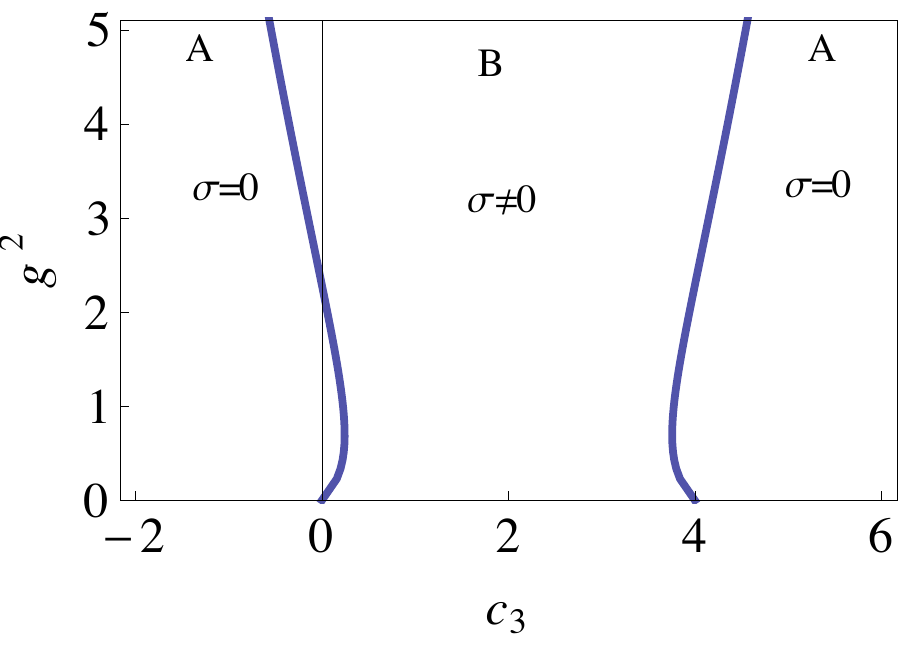}
\centering
\caption{Chiral boundaries in the parametric space i.e. $c_{3}$ vs $g^2$ for BC fermions} \label{chr}
\centering
\end{figure*}

 The action with the auxiliary fields becomes
 \begin{eqnarray}
   S_{BC}&=&\sum_{n}[\frac{1}{2}\sum_{\mu}\bar{\psi}_n\gamma_{\mu}(\psi_{n+\mu}-\psi_{n-\mu})+
   \frac{i}{2}\sum_{\mu}\bar{\psi}_n(\Gamma-\gamma_{\mu})(\psi_{n+\mu}+\psi_{n-\mu}) \nonumber \\ 
       && +\frac{N}{2g^{2}}[(\sigma(n)-m)^2+(\pi_{\Gamma}(n)-c_{3}+2)^2]\nonumber \\
       && +\bar{\psi_{n}}[\s(n)+i\Gamma\pi_{\Gamma}(n)]]\psi_{n}.
\end{eqnarray}
When the auxiliary fields are integrated out, we get back the original action given by Eq.(\ref{gnmodel}).

To find the gap equations and  chiral boundaries (i.e., the boundary between the chirally symmetric ($\sigma=0$) and broken ($\sigma\ne0$) regions), we proceed as before i.e.,  first we find the $S_{eff}$ by integrating out the fermionic fields from this equation and then use the saddle point equations $\frac{\delta S_{eff}}{\delta \sigma}=0$ and $\frac{\delta S_{eff}}{\delta \pi_{\Gamma}}=0$. The effective action after taking out the volume factor $V$  can be written as $
S_{eff}=V \bar{S}_{eff} $, where 
\be
 \bar{S}_{eff}=\frac{1}{2g^2}[(\s-m)^2+(\p-c_3+2)^2]-\int \frac{d^2k}{(2\pi)^2}\log{\logn}.~
 \ee
 Then the gap equations  are obtained  as
 \begin{eqnarray}
   \frac{(\s-m)}{g^2}&=&\int \frac{d^2k}{(2\pi)^2}\frac{2\s}{\den}~,  \label{sigg} \\
   \frac{(\p-c_3+2)}{g^2}&=&\int \frac{d^2k}{(2\pi)^2}\frac{\p+(C+D)}{\den}~;
   \ee
   where 
   \be  C & =& \sin(k_1)-\frac{1}{2}(\cos(k_1)-\cos(k_2)) ,\nonumber\\
            D &= & \sin(k_2)-\frac{1}{2}(\cos(k_2)-\cos(k_1) ).\nonumber
 \ee
  %Here we consider a single coupling constant for both the four fermi interactions,
  For the analytic calculation of the phase structure  in the parameter space we need to take $m=0$.
  % And consider that at the boundary the value of $c_{3}$ becomes $(c_{3})_c$. 
  Here  the order parameter for the phase transition is $\s$ and its value is zero or non-zero depending on the values of $c_{3}$ and $g^2$.  To get the chiral boundary  we take   $\s\to 0$ and  get the saddle-point  equations for the chiral boundaries as
  \begin{eqnarray}
   \frac{1}{g^2}&=&\int\frac{d^2k}{(2\pi)^2}  \frac{2}{\deno} \label{sigg1}, \\ 
  \frac{(\p-(c_3)_c+2)}{g^2}&=&\int \frac{d^2k}{(2\pi)^2} \frac{\p+(C+D)}{\deno},
  \end{eqnarray}
  where $(c_3)_c$ is the value of $c_3$ at the chiral boundaries.
   Using these  equations we plot the phase diagram in Fig(\ref{chr}). In this plot,  A  stands for the chiral symmetric phase and B  stands for the chiral broken phase.  The chiral critical lines  are  at two values of $c_{3}=0,4$ for zero coupling($g^2=0$) and these are the boundaries of  two minimally doubled regions. %these are our two minimal doubling region, 
    In the weak coupling limit, the regions outside this (i.e., for $c_3<0, ~c_3>4$) describe a fermionless theory as we cannot find any zero of the Dirac operator and hence  cannot cause any spontaneous chiral symmetry breaking\cite{misumi}. Thus for low couplings,  the chiral critical line  is actually the boundary between the two flavour and non-flavour phases. 
However for the values close  to $c_{3}=0,4$ and within minimally doubled regions, we find a symmetric phase at low coupling and as the coupling grows the symmetry gets broken.
Now to show that this is  a second order phase transition we show that
  the mass of $\s$ is  zero on the critical line.   The analytic value of $m_\s$ on the critical line 
%   which indicates a second order phase transition:\\ 
 \begin{align}
 \centering
   m^2_{\s}&\propto  V\frac{\delta^2 \bar{S}_{eff}}{\delta\s^2}\Bigg{\lvert}_{(c_{3})_c}  \nonumber \\
        &=V[\frac{1}{g^2}-\int\frac{2}{\den}\frac{d^2k}{(2\pi)^2} \nonumber \\
        &-\int\frac{4\s^2}{\den}\frac{d^2k}{(2\pi)^2}]\Bigg{\lvert}_{(c_{3})_c} \nonumber \\
        &=0.
\end{align} 
 The last step followed from  Eq.(\ref{sigg1}) and with  $\s=0$.  The result implies that the chiral phase transition in the the Gross-Neveu model is indeed a second order phase transition, as was shown in Ref. \cite{misumi} for the Karsten-Wilczek fermion.

%%%%%%%%%%%%%%%%%%%%%%%%%%%%%%%%%%%%%% LATTICE SIMULATION OF THE GROSS NEVEU MODEL %%%%%%%%%%%%%%%%%%%%%%%%%%%%%%%%%%%%%%%%%%%%%%%%%%%

\section{Lattice simulation of the Gross-Neveu model}
%%%%%%%%%%%%%%%%%%%%%%%%%%%%
Though there are strong coupling analyses in the literature, 
until now, the Gross-Neveu model with BC fermion has not been studied  numerically  on a lattice.   
For the lattice simulation of the Gross-Neveu model with BC fermion to study the chiral  phase transition,  we consider %$c_3=0$.
$c_3=0+\epsilon$(as at $g\approx 0$,  the exact value of $c_3=0$ does not describe minimally doubled fermion). We take a very small $\epsilon$  ($\epsilon=10^{-5}$),
%to shift the $c_3=0$ to
so that we are in  a physical two flavor region as well as close enough to the critical line to study the phase transition. 

The lattice version of the action  is given in Eq.(\ref{gnmodel}). Here, we rewrite the action explicitly in terms of the auxiliary fields and  for the chosen  parameter $c_3=0+\epsilon$ as 
   \begin{equation}
     S=\bar{\psi_{i}}M_{ij}\psi_{j}+{\frac{N}{2g^{2}}}(\sigma^{2}+\pi_{\Gamma}^{2}),
   \end{equation}
   where the auxiliary fields
   \begin{eqnarray}
  \s=-\frac{g^2}{N}(\bar{\psi}\psi), \nonumber\\
  \p=-\frac{g^2}{N}(\bar{\psi}i\Gamma\psi)
\end{eqnarray}
are defined  in the dual lattice sites $\tilde{x}$ surrounding the direct lattice site $x$ \cite{hands2} and 
 \be 
 M_{ij}=D_{ij}+\frac{1}{4}\sum_{\langle x,\tilde{x}\rangle}{(\sigma+i\pi_{\Gamma}\Gamma)},
 \ee   where $D_{ij}$ is the BC Dirac operator:
 \be 
 D_{ij}=\frac{1}{2}\gamma_\mu(\delta_{j,i+\mu}-\delta_{j,i-\mu})+\frac{i}{2}(\Gamma-\gamma_\mu)(\delta_{j,i+\mu}+\delta_{j,i-\mu})-((2-\epsilon)i\Gamma-m)\delta_{i,j}.
 \ee
 Because of species doubling,  two dimensional $N$ flavors of fermions correspond to $N_{f}=2N$ continuum flavors.
Since $M$ is a complex matrix, we work with (${M}^{\dagger}M$) to make it real and positive definite and  integrate out the fermion fields by the pseudofermion method.
 Since  the {\CB} action describes two flavors, the number of flavors becomes double i.e., $N_f=4$ for an action with  ($M^\dagger M$).
%  and for $N_{f}=4$ flavours,
 With pseudofermions the action becomes
 \begin{equation}
  S=\phi^{\dagger}(M^{\dagger}M)^{-1}\phi+{\frac{1}{g^{2}}}(\sigma^{2}+\pi_{\Gamma}^{2}).  
\end{equation}
 We simulate our model by the  HMC method and evaluate the order
   parameter for the chiral  phase transition $\la{\s}\ra$ as a function of coupling constant. We use  a single noise vector to estimate the condensate.
  \begin{eqnarray}
    \la\bar{\psi}{\psi}\ra&=&-\la{TrM^{-1}}\ra \\
    \la\s\ra&=&-\beta{\la{\bar{\psi}{\psi}}}\ra \\
    \mbox{where}, \beta&=&\frac{1}{g^2}.  \nonumber 
  \end{eqnarray}
  The  configurations are generated  by considering stepsize($\triangle {t}$)=0.1 in the leapfrog method and  ten %number of
   steps per trajectory in the molecular dynamics chain. We do not use any preconditioning during the simulation.
    The first 500 ensembles are rejected for thermalization and  data  are collected from the next 16000 ensembles by varying the $\beta$. For stability of the code, we cannot take exact  massless fermion.  So we expect that the lattice results for light fermions  might have slight differences from the analytic results for $m=0$  presented in the previous section. 
   Here we consider only  (very) light fermions   so
   %so \DC{the chiral symmetry breaking due to the mass term is very small relative to the coupling constant considered for this study 
   %(the maximum value of $m$ we have considered is 0.03 while the  value of $g$ corresponding to $\beta=1.6$ is  0.79).}
   that the chiral symmetry of the action is still approximately valid and deviation from $m=0$ results is minimal. 
   For very low mass,  the time for a single update  increases rapidly  with increasing $\beta$
    and the number of conjugate gradient iterations per update is around 2000 to 5000 for each case.
  The order parameter  $\la\s\ra$ is plotted against $\beta$ for  three masses on a  $32\times32$ lattice in Fig.\ref{svbeta}. The plot indicates   that a second order phase transition exists.
  \begin{figure*}
  \includegraphics[width=14cm]{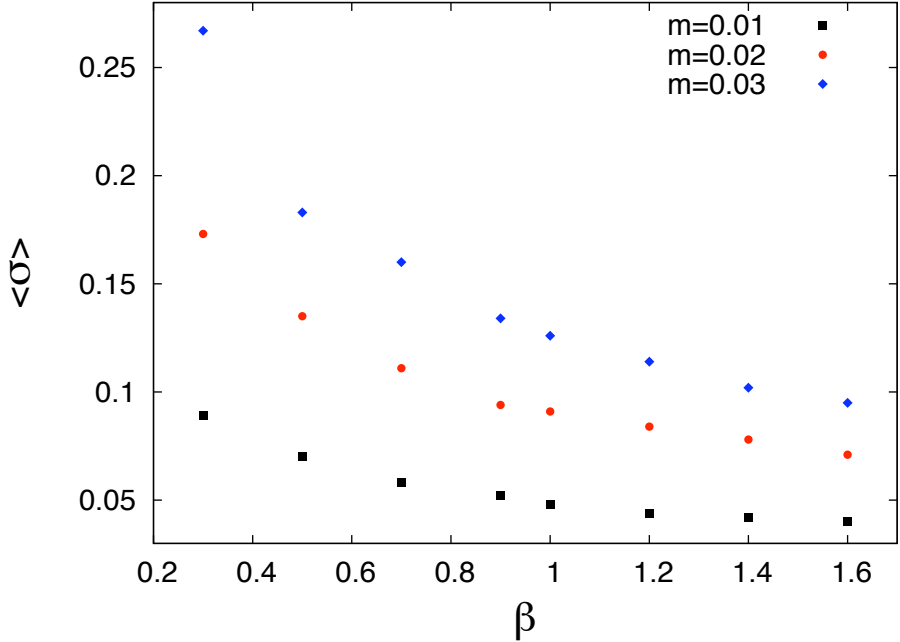}
  \centering
\caption{${\la\sigma}\ra$ vs $\beta$ of $m=0.01,0.02$ \& $0.03$ for Gross-Neveu model with BC fermions in a 32$\times$32 lattice.\label{svbeta} }
\centering
\end{figure*}
 The numerical results on the lattice show that at  low coupling(i.e.$\beta$ large), the order parameter $\s$ is very small and follows the perturbative RG flow with the coupling constant. But as the coupling becomes stronger, $\s$ grows to a nonzero value indicating a phase transition. 
 %Since $m\ne0$ in the lattice simulations, as the value of $m$ decreases, the critical coupling constant approaches towards the analytic result $\beta_c\approx 0.4$ for massless fermion.  
From the analytic calculation we have the critical value of $\beta$ as $\beta_c\approx 0.4$. In  the lattice calculation, we have non-zero fermion mass, so the numerical results for the critical couplings  for different masses are not the same as  the analytical result, but as the masses are small, they are close to $\beta_c$. As $m\to0$, one expects to reproduce the analytical value of $\beta_c$ for  the massless fermion. 
  Though the numerical results are not conclusive about the order of the phase transition but 
  note that the order parameter $\s$ varies smoothly near the phase transition supporting the analytical prediction of a second order phase transition.
  For any study of critical behavior, the knowledge of finite size effects is very important.
  In Fig.\ref{finVol}, we have shown the finite volume effects for $m=0.03$ for three different lattice volumes of $20\times 20$, $32\times 32$ and $40\times 40$.  The plot shows that the finite volume effect is very small and  negligible for  the lattice volumes considered in this work.

\begin{figure*} 
\centering
%\begin{minipage}[b]{0.45\linewidth}
\includegraphics[width=14cm]{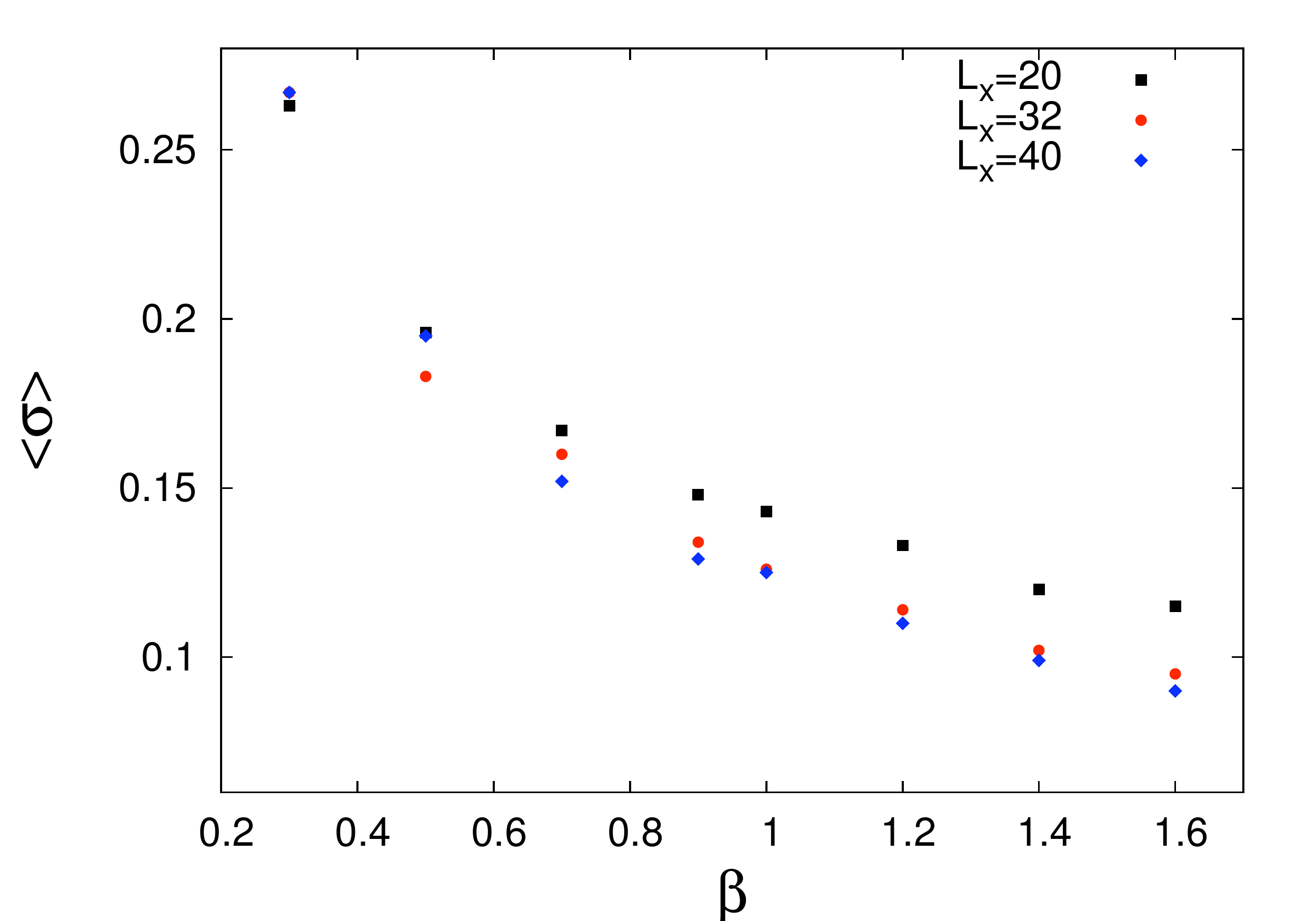}
\caption{Finite volume effects of $\la\sigma\ra$ vs $\beta$ for $ m=0.03$  of three different lattice sizes $20\times 20,~32\times 32$, and $40\times 40$.\label{finVol}} 
\end{figure*}

  In the analytic calculation, only two flavors were considered as required by minimally doubled lattice formulation (BC action), but numerically  we  simulate with four light flavors as we double the number of flavors  to have a real and positive definite  action. Our numerical results are in good agreement with the analytic results and show that due to the four fermion interactions, at a high coupling limit the chiral symmetry is dynamically broken. 
%%%%%%%%%%%%%%
\section{Summary}
%%%%%%%%%%%%
In this work, we have studied the Gross-Neveu model with  the minimally doubled fermion action proposed by Creutz and Borici.   First we have  studied the chiral phase structure of the GN model  analytically which predicts  a second order phase transition from  the symmetric to broken chiral phase.  Then we have studied the model with the  HMC algorithm. Our numerical simulations also agree with the analytic analysis and the order parameter $\langle\s\rangle$ plotted against $\beta=1/g^2$ shows a second order phase transition.

%%%%%%%%%%%%%%%%%%%%%%%%%%%%%%%%%%%%%%%%%


\begin{thebibliography}{99}
%%%%%%%%%%%%%%%%%%%%%%%%%%%%%%%%%%%%%%%%%%%%%%%%%%%%%%%%%%%%%%%%%%%%%%%%%%%%%%%

\bibitem{K} L.H. Karsten, Phys. Lett. B  {\bf 104},315 (1981).
\bibitem{W} F. Wilczek, Phys. Rev. Lett. {\bf 59},2397 (1987).
\bibitem{creutz} M. Creutz, JHEP {\bf 0804}, 017(2008).
\bibitem{borici} A. Borici, Phys. rev. D. {\bf 78}, 074504 (2008).
\bibitem{bedaque} P. F. Bedaque, M. I. Buchoff, B.C. Tiburzi, A. Walker-Loud, Phys. Lett. B. {\bf 662},449 (2008).
\bibitem{creutz2} M. Creutz, Pos LAT2008, 080 (2008).
\bibitem{capitani} S. Capitani, M. Creutz, J. Weber, H. Wittig, JHEP {\bf 1009}, 027 (2010).
\bibitem{dc} D. Chakrabarti, S. J. Hands, A. Rago, JHEP {\bf 0906}, 060 (2009).
\bibitem{CKM} M. Creutz, T. Kimura, T. Misumi, Phys. Rev. D.{\bf 83}, 094506 (2011).
\bibitem{misumi} T. Misumi, JHEP  {\bf 1208}, 068 (2012).
\bibitem{korzec} T. Korzec, F. Knechtli, U. Wolff, B. Leder, PoS LAT2005, 267 (2006).
\bibitem{hands} S. J. Hands, C. Strouthos, Phys. Rev. B{\bf 78}, 165423 (2008), W. Armour, S. Hands, C. Strouthos, Phys. Rev. B{\bf 81},  125105(2010).
\bibitem{araki} Y. Araki,  PoS LAT2011, 054 (2011); Phys. Rev B {\bf 85}, 125436 (2012).
\bibitem{basak} S. Basak and A. K. De, Phys. lett. B {\bf 430}, 320 (1998).
\bibitem{kimura} T. Kimura, S. Komatsu, T. Misumi, T. Noumi, S. Torii, S. Aoki, JHEP {\bf 1201}, 048 (2012).
\bibitem{hands2} S. J. Hands, A. Koci\'c, J. B. Kogut, Nucl. Phys. B{\bf 390}, 355 (1993), Ann. Phys. {\bf 224}, 29 (1993).


%%%%%%%%%%%%%%%%%%%%%%%%%%%%%%%%%%%%%%%%%
\end{thebibliography}
\end{document}